\documentstyle[psfig]{mn}
\newcommand{\HeI} {He\,{\sc i}}
\newcommand{\HeII} {He\,{\sc ii}}

\newcommand{\NIII} {N\,{\sc iii}}

\newcommand{\CIII} {C\,{\sc iii}}
\newcommand{\CIV} {C\,{\sc iv}}

\newcommand{\OII} {O\,{\sc ii}}

\newcommand{\SiIII} {Si\,{\sc iii}}

\newcommand{\SiIV} {Si\,{\sc iv}}
\newcommand{\MgII} {Mg\,{\sc ii}\,}

\newcommand{\CaII} {Ca\,{\sc ii}\,}

\newcommand{\NaI} {Na\,{\sc i}\,}
\newcommand{\teff}{$T_{\rm eff}$}
\newcommand{\g} {$\log g$}
\newcommand{\Mv} {$M_{\rm v}$}
\newcommand{\eps} {$\epsilon$}

\newcommand{\lMp} {log \.M}

\newcommand{\Hg} {H${\rm \gamma}$}
\newcommand{\Hb} {H${\rm \beta}$}
\newcommand{\Ha} {H${\rm \alpha}$}
\newcommand{\Rsun} {$R_{\odot}$}
\newcommand{\Msun} {$M_{\odot}$}
\newcommand{\R} {$R/R_{\odot}$}
\newcommand{\M} {$M/M_{\odot}$}

\newcommand{\lL} {$\log (L/L_{\odot})$}

\newcommand{\lR} {$\log (R/R_{\odot})$}
\newcommand{\lM} {$\log (M/M_{\odot})$}

\title[The spectral variations of the O-type runaway supergiant HD 188209]
{The spectral variations of the O-type runaway supergiant HD 188209
\author[G. Israelian et al.]
{G.~Israelian,$^1$
 A.~Herrero,$^1$ 
 F.~Musaev,$^2$
 A.~Kaufer,$^3$
 A.~Galeev$^{4,2},$
 G.~Galazutdinov$^2$\cr and E.~Santolaya-Rey$^1$\\
 $^1$Insituto de Astrofisica de Canarias, E-38200 La Laguna, Tenerife,
 Canary Islands, Spain\\
 $^2$Special Astrophysical Observatory of the Russian AS, Nizhnij Arkhyz
 357147, Russia\\
 $^3$ ESO, Karl-Schwarzschild-Str. 2, D-85748, Garching, Germany\\
 $^4$ Department of Astronomy, Kazan State University, Kazan, Kremlevskaja Str.,
 420008, Russia}}
\date{}
\pagerange{\pageref{firstpage}--\pageref{lastpage}}
\begin{document}
\label{firstpage}
\maketitle
\begin{abstract}
{
We report spectral time series of the late O-type runaway supergiant 
HD 188209. Radial velocity variations of photospheric absorption  
lines with a possible quasi-period $\sim$ 6.4 days have been detected in 
high-resolution echelle spectra. 
Night-to-night variations in the 
position and strength of the central emission reversal of the 
\Ha~profile occuring over 
ill-defined time-scales have been observed. The fundamental parameters 
of the star have been derived using state-of-the-art plane--parallel and 
unified non-LTE model atmospheres, these last including the mass-loss rate. 
The derived helium abundance is moderately enhanced with respect to solar, 
and the stellar masses are lower than those predicted by the evolutionary models.
The binary nature of this star is not suggested either from {\it Hipparcos\/}
photometry or from radial velocity curves.
} 
\end{abstract}


\section{Introduction}

Runaway O stars have been defined 
as a group by Blaauw (1961), who introduced the term {\it runaway} 
to describe the space motions of AE Aur and $\mu$ Col.
Blaauw (1961) has also suggested that
such stars were ejected in the breakup of binary systems in 
supernova explosions by their companions. In later evolutionary
stages, the initial secondary appears as a most massive star
and transfers matter to the compact companion (the initial
primary) making the system appear as a massive X-ray binary
(van den Heuvel 1976). 
Given the possibility of the binary nature of runaway stars,
it appears to be an important task to measure the radial velocity (RV)
variations of the photospheric lines. Systematic searches 
for RV variations have been made in order to assess the
binary frequency of O stars (e.g. Garmany, Conti \& Massey 1980; 
Stone 1982, Gies 1987). In many cases the amplitude of RV variations is 
quite large, and the additional presence of a clear periodicity 
immediately suggests a binary nature for the system. 
However, there are stars which show more complicated RV curves, 
and the interpretation of their spectral variability is not 
straightforward. HD 188209 (O9.5Iab) is one of those objects. 
Garmany et al. (1980) have concluded from three spectra that this 
star is probably not a binary, and that the RV variations must be 
attributed to atmospheric motions. This conclusion was supported
by Musaev \& Chentsov (1988). However, based on 21 measurements 
Stone (1982) has concluded that HD 188209 can be considered
as a spectroscopic  binary with  a period 57 days and small 
semiamplitude. More recently, Fullerton, Gies \& Bolton (1996) 
included HD 188209 in a large sample of stars investigated on
the presence of line profile variability (LPV) and found 
LPVs only in \HeI~5876 \AA. However, they did not flag 
HD 188209 as a velocity variable (their Table 10). 

The binarity of many
O supergiants has been proposed recently by Thaller (1997).
The fact that binaries have a higher incidence and an H$\alpha$
emission strength in post-MS stages may indicate that wind
interactions are a common source of emission in massive stars.
In other words, even in cases where RV measurements are not
available, the presence of \Ha~emission in
a spectrum could be linked with colliding winds. One needs to study
orbital phase variations in the \Ha~profile in order
to be sure that the latter is due to colliding winds instead of
some other mechanism. Note that HD 188209 is an X-ray source
detected by {\it ROSAT} (Berghoefer, Schmitt \& Cassinelli 1996). 

In this paper we focus on the high-resolution spectroscopic data
of HD 188209. Our observations can possibly account for 
the small semi-amplitudes and eccentric orbits of this binary 
candidate since they have been accumulated at different 
periods over a long baseline. 

\section{Observations}

The observations have been carried out in different runs (Table 1) using 
the the Coud\'{e} Echelle Spectrometer (Musaev 1993) at the 1-m telescope
of the Special Astrophysical Observatory of the Russian Academy of
Science. Most of the spectra have a signal-to-noise ratio
S/N$\ge$100 per resolution element, and an average resolution 
$R = $ 40000 in the wavelength region 4400--7000 \AA. Preliminary 
reduction of the echelle spectra CCD images was made using 
the {\sc dech} code (Galazutdinov 1992), which allows
the flat-field division, bias/background subtraction, one-dimensional 
spectrum extraction from two-dimensional images, 
excision of cosmic-ray features, spectrum addition, 
correction for diffuse light, etc. Numerous bias, flat-field
have been obtained every night. Each image was subject to a
bias-frame subtraction and flat-field division using nightly
means.  Comparison exposures of a Th-Ar lamp were taken for each 
stellar spectrum. The control measurements of interstellar \CaII~
and \NaI~D lines revealed a small scatter of 0.8 $~{\rm km}~{\rm s}^{-1}$ 
(1 $\sigma$). However, the 1 $\sigma$ dispersion of the velocity
of the DIB was 1.5 $~{\rm km}~{\rm s}^{-1}$. These interstellar
lines have been used to align all the spectra in the time series
accurately. As an indicator of the overall
precision of our measurements we have adopted 1 $\sigma$ 
dispersion 1.5 $~{\rm km}~{\rm s}^{-1}$ of the velocity of the DIB. 
All stellar absorption lines exhibited variations about their
respective mean velocities at least 2-3 times the dispersion 
of the DIB velocities. The mean rms obtained from different 
dispersion curves was at least 0.003 \AA. The Coud\'{e} Echelle 
Spectrograph was not a subject to mechanical and/or thermal instabilities.  

 We used a  580$\times$530 (pixel size 24 $\times$ 18 $\mu$m)
CCD camera in all runs except 8 and 9. The last run was carried out 
using the  Coud\'{e} Echelle Spectrograph (Musaev, 1999) at the 2-m 
telescope located in Terskol (North Caucasus, Russia). The CCD used in
last two runs had a larger matrix (WI 1242$\times$1152 pixel with
pixel size 22.5 $\times$ 22.5 $\mu$m) allowing a coverage in a 
single exposure of the region $\sim$ 3500--10100 \AA\  with almost 
the same resolution. 
 
\def\baselinestretch{1}

\begin{table}
\caption{Journal of Observations. The second column gives the universal 
plus exposure time and in third column is the spectral region. 
The S/N ratio is given at 6600 \AA.}
\begin{flushleft}
\begin{tabular}{lllrr}
\hline\noalign{\smallskip}
Run & Date & UT + EXP & Region & S/N \\
    &(dd.mm.yy)&      & (\AA)   &    \\
\noalign{\smallskip}
\hline\noalign{\smallskip}

1 & 14.06.93 &  20h 35m+30m & 4800--7500  & 100 \\
  & 07.07.93 &  20h 40m+30m & 4420--6710  & 100 \\
  & 08.07.93 &  19h 25m+20m & 4350--4510  & 200 \\
  &          &              & 6540--6720  & \\
2 & 04.09.93 &  21h 53m+30m & 4350--4510  & 200 \\
  &          &              & 6540--6720  & \\
  & 05.09.93 &  23h 55m+20m & 4350--4510  & 200 \\
  &          &              & 6540--6720  & \\
  & 06.09.93 &  20h 40m+20m & 4350--4510  & 200\\
  &          &              & 6540--6720  & \\
  & 07.09.93 &  20h 07m+45m & 4350--4510  & 200 \\
  &          &              & 6540--6720  & \\
  & 09.09.93 &  22h 40m+45m & 4350--4510  & 200 \\
  &          &              & 6540--6720  & \\
3 & 07.10.93 &  17h 46m+60m & 4370--6720  & 150 \\
  & 08.10.93 &  17h 35m+60m & 4370--6720  & 150 \\
  & 09.10.93 &  17h 40m+60m & 4370--6720  & 150 \\
  & 10.10.93 &  18h 32m+60m & 4370--6720  & 150 \\
4 & 31.10.93 &  16h 22m+30m & 4330--4520  & 200 \\
  &          &              & 6540--6720  & \\
  & 01.11.93 &  20h 32m+30m & 4330--4520  & 200 \\
  &          &              & 6540--6720  & \\
  & 02.11.93 &  16h 18m+45m & 4400--6800  & 100 \\
5 & 24.11.93 &  17h 56m+45m & 4420--7000  & 100 \\
  & 25.11.93 &  17h 02m+60m & 4420--7000  & 150 \\
  & 27.11.93 &  19h 15m+60m & 4420--7000  & 150 \\
  & 28.11.93 &  16h 50m+60m & 4420--7000  & 150 \\
  & 29.11.93 &  16h 41m+60m & 4420--7000  & 150 \\
  & 30.11.93 &  15h 50m+60m & 4420--7000  & 150 \\
  & 02.12.93 &  16h 22m+60m & 4420--7000  & 150 \\
  & 05.12.93 &  17h 08m+60m & 4420--7000  & 150 \\
6 & 16.09.94 &  17h 27m+60m & 4420--7000  & 150 \\
  & 17.09.94 &  17h 26m+60m & 4420--7000  & 150 \\
  & 18.09.94 &  17h 18m+60m & 4420--7000  & 150 \\
  & 19.09.94 &  17h 36m+60m & 4420--7000  & 150 \\
  & 20.09.94 &  21h 21m+60m & 4420--7000  & 150 \\
  & 21.09.94 &  18h 01m+60m & 4420--7000  & 150 \\
7 & 13.10.94 &  17h 00m+60m & 4350--6710  & 150 \\
  & 15.10.94 &  16h 40m+60m & 4350--6710  & 150 \\
  & 16.10.94 &  20h 20m+60m & 4350--6710  & 150 \\
  & 17.10.94 &  18h 01m+60m & 4350--6710  & 150 \\
  & 25.10.94 &  18h 00m+60m & 4350--6710  & 150 \\
  & 27.10.94 &  19h 06m+40m & 4350--6710  & 100 \\
  & 29.10.94 &  16h 36m+40m & 4350--6710  & 100 \\
8 & 15.05.97  &  00h 35m+45m & 3380--10060  & 200 \\
  & 15.05.97  &  21h 56m+60m & 3380--10060  & 250 \\
  & 16.05.97  &  21h 09m+60m & 3380--10060  & 250 \\
  & 30.08.97  &  19h 03m+30m & 4420--6800  &  100 \\
9 & 08.03.98   &  02h40m+30m  & 3560--10060  & 200 \\
  & 09.03.98   &  01h50m+45m  & 3560--10060  & 250 \\
  & 10.03.98  &  02h02m+45m  & 3560--10060  & 250 \\
  & 11.03.98  &  01h32m+20m  & 3560--10060  & 200 \\
\noalign{\smallskip}
\hline
\end{tabular}
\end{flushleft}
\end{table}

\begin{figure*}
\vbox{\psfig{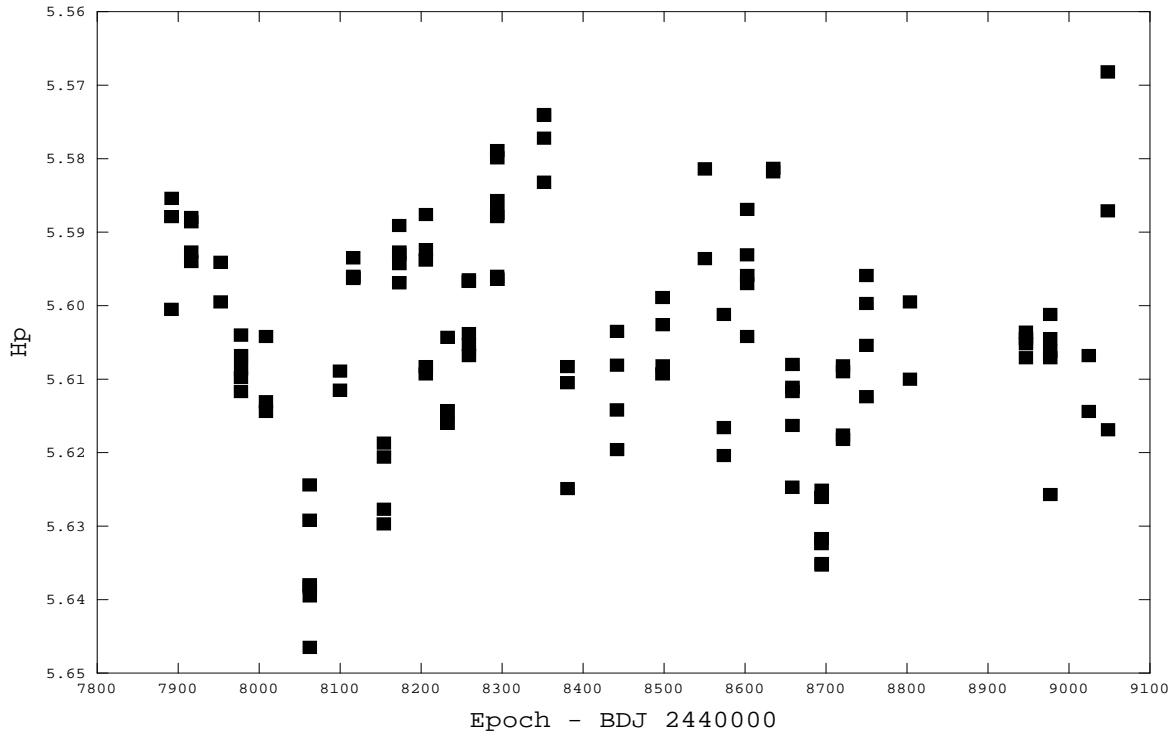}}
\caption[]{ {\it Hipparcos} photometry of HD 188\,209.}
\end{figure*}

\def\baselinestretch{2}

\section{Photometry}

The photometry of HD 188209 obtained with
{\it Hipparcos} (ESA 1997) is presented in Figure 1. 
The approximate response curve for the Hp$_{\rm dc}$ passband (see
van Leeuwen et al. 1997) is very extended with a maximum at
$\sim$ 4500 \AA. Our observations did not reveal a strong
variability in the equivalent widths of the
strongest lines in the spectrum of HD 188209, which suggests 
that the variations detected by {\it Hipparcos}
in the Hp$_{\rm dc}$ band are due to changes in the 
continuum flux. The  mean  value of Hp$_{\rm dc}$=5.605 has a 
standard deviation 0.0155. The level of photometric variability is 
significant and cannot be ascribed to standard errors in the
Hp$_{\rm dc}$ magnitude of the order of 0.005 (van Leeuwen et al. 1997).
Tests on the periodicity of the photometric variations were
performed but no convincing period has been found. However,  
considering the large gaps in different {\it Hipparcos} measurements
we cannot definitely rule out short-term photometric periodicity.

\section{Fundamental Parameters of HD~188209}

\subsection{Plane--parallel analysis}

To analyse the spectrum of HD 188209 we first determine the 
rotational velocity from the 
width of 11 metal lines of C, N, O, Si, Mg and Ca, adopting
a Gaussian instrumental profile with a FWHM of 0.13 \AA.
The resulting value was a projected rotational velocity
of 82.0 $\pm$ 8.5 km~s$^{-1}$, in good agreement with 
the value of 87$~{\rm km}~{\rm s}^{-1}$ reported by Penny (1996),
and between those given by Conti and Ebbets (1977) and
Howarth et al. (1997) who give 70 and 92$~{\rm km}~{\rm s}^{-1}$,
respectively. 

The method followed in determining the stellar parameters from the
spectrum using NLTE, plane--parallel hydrostatic model atmospheres
has been described in detail by Herrero et al. (1992, and 
references therein). Briefly, we determine, at a fixed helium 
abundance, the gravity that best fits the different  
H and He profiles at a given temperature for a set of temperatures.
If the abundance is right, the lines in the \teff --\g~ diagram
will ideally cross at a point, giving the stellar \teff~ and \g. Usually,
they form an intersection region, whose central point is taken as
giving the stellar parameters, and whose limits give the adopted error.
If the lines do not cross at any point, the helium abundance is changed.
The helium abundance giving the smaller intersection region for all
profiles is the one selected. The center of the intersection region is 
taken again as that giving the stellar parameters.

Recently, McErlean, Lennon \& Dufton~\cite{mc98} and Smith \& Howarth~ 
\cite{sh98} have shown that different \HeI~ lines give different
helium abundances in the region of the \teff -- \g~ diagram
occupied by HD 188209. They attribute this to the effect
of microturbulence and show that a value of around 10 km s$^{-1}$
is appropriate for bringing most of the \HeI~ lines into agreement.
 Thus, we adopt 
this value for HD 188209 and carry out the analysis in the way 
described above.

\def\baselinestretch{1}

\begin{figure*}
\label{spfit}
\psfig{{figure=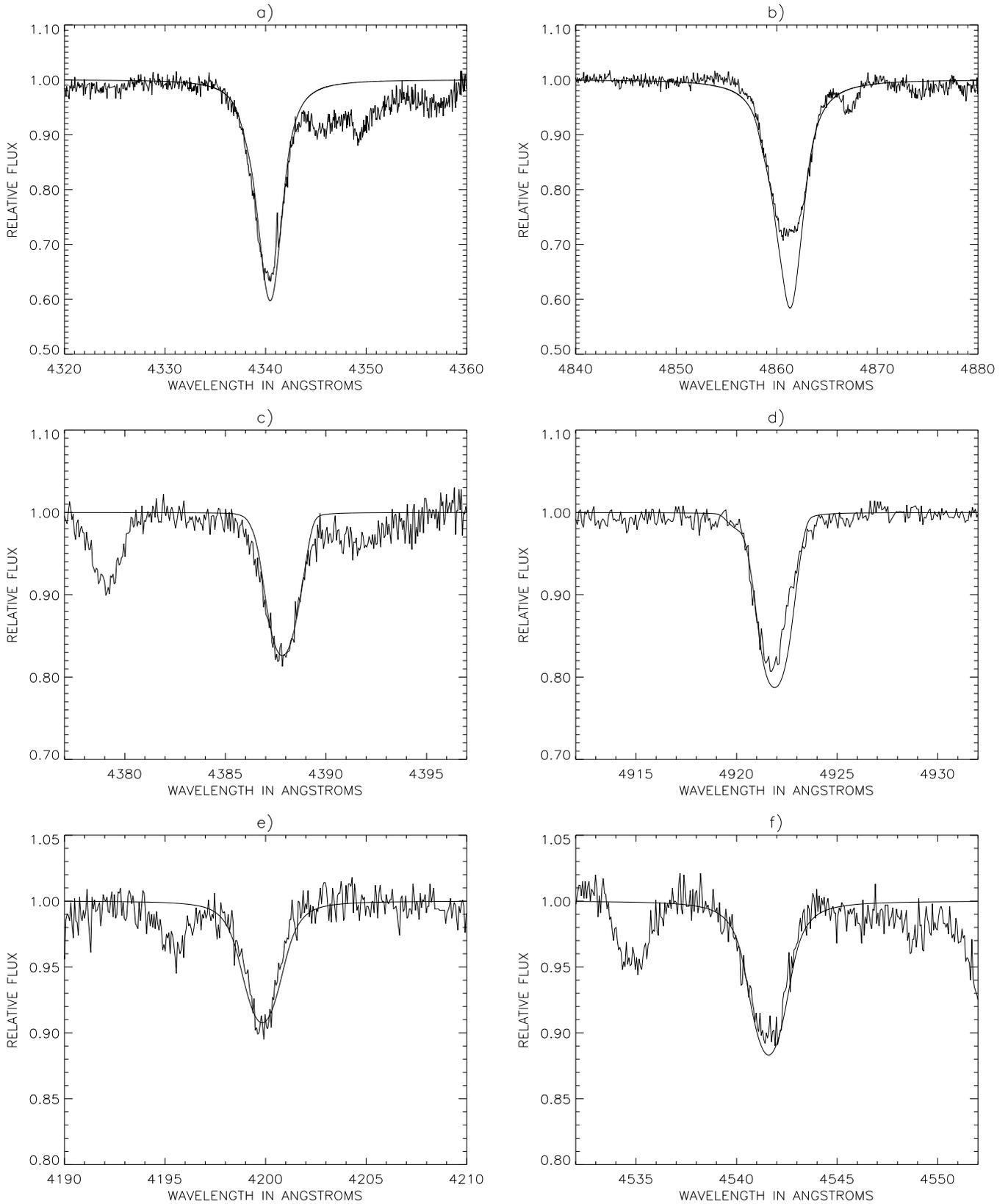,width=18.5cm,height=22cm}}
\caption[] {The fit to the HD 188209 lines using unified models.
a) \Hg ; b) \HeI~ 4471; c) \HeI~ 4387;
d) \HeI~ 4922; e) \HeII~ 4200, and f) \HeII~ 4541 \AA (see text for details).}
\end{figure*}

\begin{figure*}
\label{spfit}
\psfig{{figure=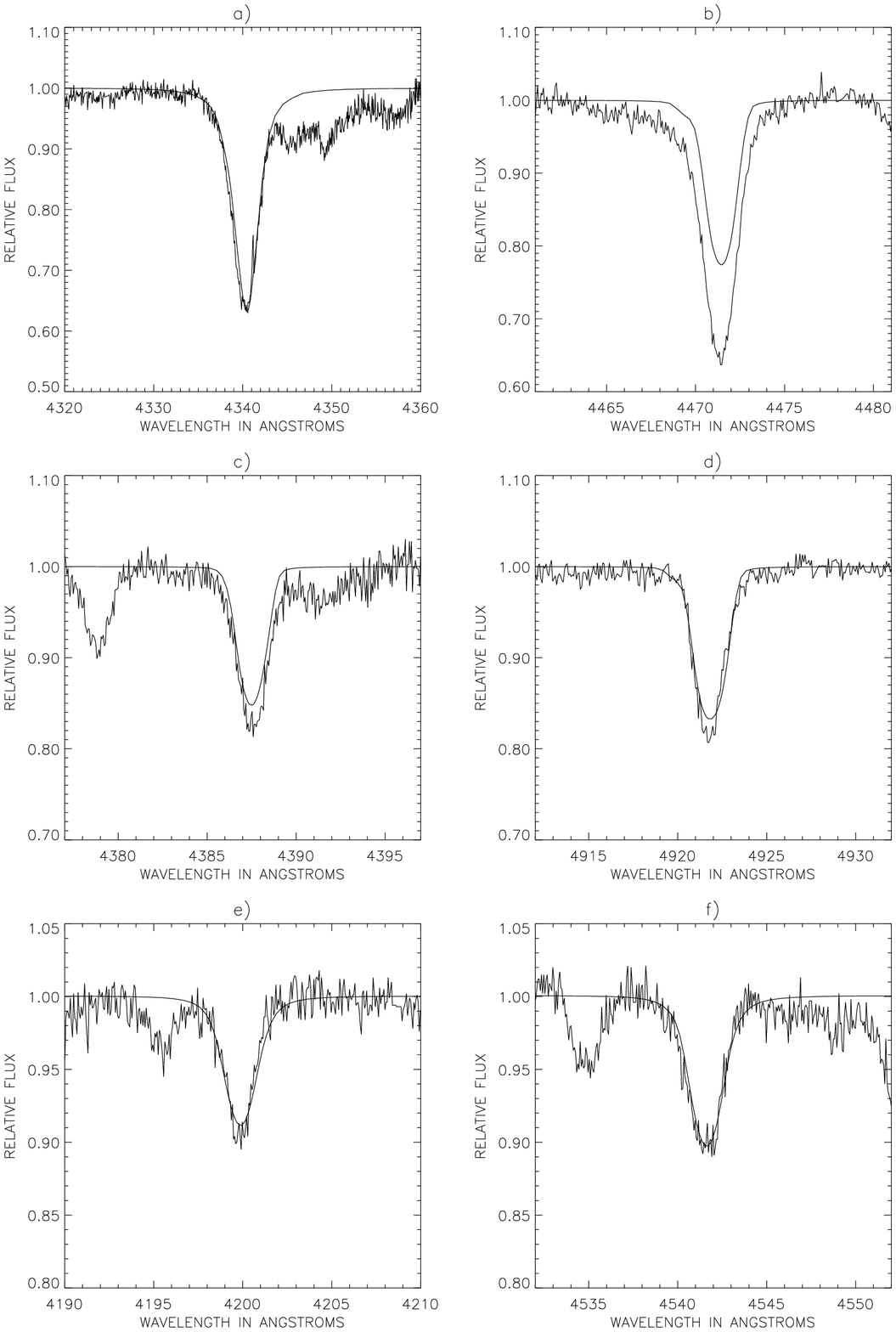}}
\caption[] {The fit to the HD 188209 lines using unified models.
a) \Hg ; b) \HeI~ 4471; c) \HeI~ 4387;
d) \HeI~ 4922; e) \HeII~ 4200, and f) \HeII~ 4541 \AA (see text for details).}
\end{figure*}

\def\baselinestretch{2}

With this method we have determined the stellar parameters of HD 188209.
We obtained \teff= 31\,500 K $\pm ^{1000}_{500}$, \g= 3.0 $\pm 0.1$
(uncorrected for centrifugal force; because of the low rotational velocity
we neglect this small correction here) and \eps= 0.12 $\pm 0.03$
(the abundance of helium
with respect to the total abundance of hydrogen plus helium, by number;
the solar abundance is \eps= 0.09). The final fits for the H and He
lines are shown in Figure 2.

The larger errors towards higher temperatures is due to a small difference
in the two spectrograms available for \HeII~4541 \AA. Using the 
second one, we would have obtained a \teff~ of 32\,000 K, all other
parameters remaining the same. For this reason, we have enlarged
the error bar in this direction. Remember that the errors given are
formal errors, in the sense that they express the uncertainty
in the fit using the models described above.
 
As  has been show by Herrero \cite{h94} the metal opacity in the
UV (line blocking) could also affect the value of the parameters 
determined. However, at the relatively low temperature of HD 188209
the effect would be minor, moving \teff~ towards higher temperatures
within the error box.

With the stellar parameters given above we can determine the
radius, luminosity and mass of HD 188209 as described in Herrero 
et al. (1992). For \Mv = $-$6.0 mag ($M_{\rm bol}$ = $-$9.0 mag) 
given by Howarth \& Prinja (1989) we obtain \R = 20.9, \lL = 5.59 
and \M = 16.6. The errors are again as in Herrero et al. (1992): 
$\pm$0.06 in \lR, $\pm$0.16 in \lL~ and $\pm$0.22 in \lM.

HD 188209 does not formally show the helium discrepancy, as the solar
helium abundance is within the error bars. However, it shows
the mass discrepancy: the mass derived from the plane--parallel
spectroscopic analysis is, even including the error bars, much
lower than the one derived from the evolutionary tracks
from Schaller et al. \cite{sch92}.

\subsection{Unified model analysis}

After having the parameters from the plane--parallel analysis, we
can try to use a spherical, non-hydrostatic model
atmosphere in order to improve the already derived 
parameters and also to obtain the mass-loss rate. In a supergiant
like HD 188209 this can have an important impact on the
final parameters. Usually, it is also assumed that this will
contribute to the reduction in the mass discrepancy.

The unified code we use is that recently developed by Santolaya--Rey,
Puls \& Herrero \cite{sph97}. The reader will find all the details 
therein, but for our present purposes we mention that the code uses
spherical geometry, with a $\beta$--velocity field, and treats
the wind and the photosphere in a unified way. It also makes
use of the NLTE-Hopf functions. Stark broadening is included in the
formal solution, and the model atoms are the same as in the
plane--parallel case (slightly adapted for the new program).

We begin by estimating the mass-loss rate from the \Ha~profile,
and then, with this mass-loss rate, we try again to find  the
best gravity (from the \Hg~wings), effective temperature 
and He abundance (from the He ionization equilibrium).
With the new parameters, we again try to fit the \Ha~ by 
varying the mass-loss rate, and so on. In the whole process 
we take the wind terminal velocity from Haser (1995), who
gives 1700 km s$^{-1}$. Note that Howarth et al. (1997), give a
similar value of 1650 km s$^{-1}$. 

We have used for the analysis the same lines as in Fig. 2, but have
included \HeI~4471 \AA\ instead of \Hb (that improved as does \Hg)
to show the dilution effect (see below). 
In Fig. 3 we show the fit of the \Hg, \HeI~and \HeII~lines.
The adopted parameters are now \teff = 31\,500 K, \g = 3.00
and \eps= 0.12, i.e. the same parameters as in the plane--parallel case,
even for the gravity (again adopting a microturbulent velocity of 10 
km s$^{-1}$). Thus, the unified models do not contribute
in this case to changes in the mass discrepancy
found with the plane--parallel models (nor in the He abundance).
We can also see in Fig. 3 
that the \HeI~4471 line shows the well known dilution effect
mentioned by Voels et al. \cite{vo89}, although the other He lines fit 
perfectly well. This effect merits an explanation.

The fitting of the \Ha~profile needed 
to derive the mass-loss rate cannot be done properly.
The profile is highly variable, and we have adopted a qualitative
approach: we have simply tried to give upper and lower limits to
the mass-loss rate. As an example, we illustrate the procedure
in Fig. 4, where we show one of the profiles with
various mass-loss rates for the stellar parameters given above.
The theoretical mass-loss rate values shown in Fig. 4
are \lMp = $-$5.70, $-$5.80 and $-$5.90 (with \.M in \Msun /yr). 
This situation is
similar to that found by Herrero et al. \cite{h95} for Cygnus X--1.
The profile shown cannot be adopted as either  an average or a 
representative one, as the profile varies a lot. The figure is only
for illustrative purposes. An average logarithmic mass-loss rate for 
HD~188209 would be between $-$6.0 and $-$5.7. These values are 
also in agreement with the profiles in Fig. 3, where
the adopted mass-loss rate was \lMp =$-$5.80 
(corresponding to 1.58 10$^{-6}$ \Msun /yr). We should point out
that other values of the mass-loss rate give worse
fits to the \Hg~and \HeI~lines, by strongly modifying the
line cores.

\section{Radial velocity variations of the absorption lines}

\def\baselinestretch{1}

\begin{figure}
\label{halfa}
\vbox{\psfig{figure=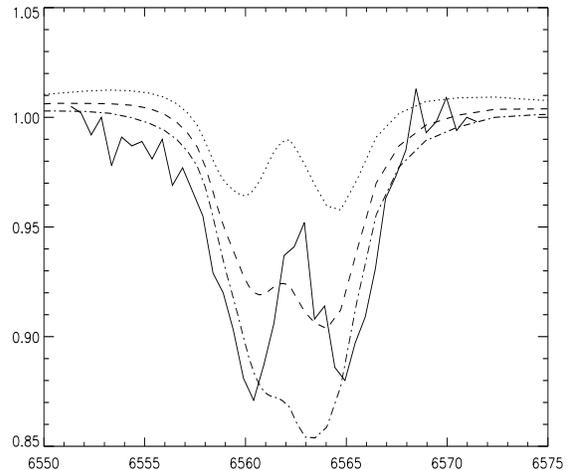,width=8.0cm,height=7.0cm}}
\caption[]{The \Ha~profiles of HD 188209 observed on 1993 
September 7 enclosed by
different theoretical profiles calculated for \lMp = $-$5.70, $-$5.80 and
$-$5.90 (dotted, dashed and dash--dotted lines, respectively)
for the parameters given in text. The selected observed 
\Ha~profile is neither an average nor 
representative, and the figure merely illustrates the approach we
have followed.}
\end{figure}

\begin{figure}
\vbox{\psfig{file=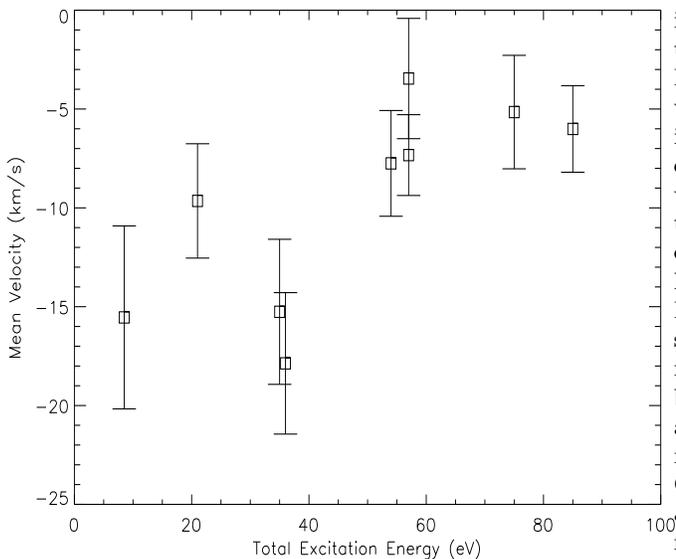,width=9.5cm,height=8.0cm,angle=90}}
\caption[]{Mean radial velocities of different groups of lines
versus TEE. A linear fit (0.1527,-17.05) provides a non-parametric 
correlation coefficient 0.73.}
\end{figure}

\begin{figure}
\vbox{\psfig{file=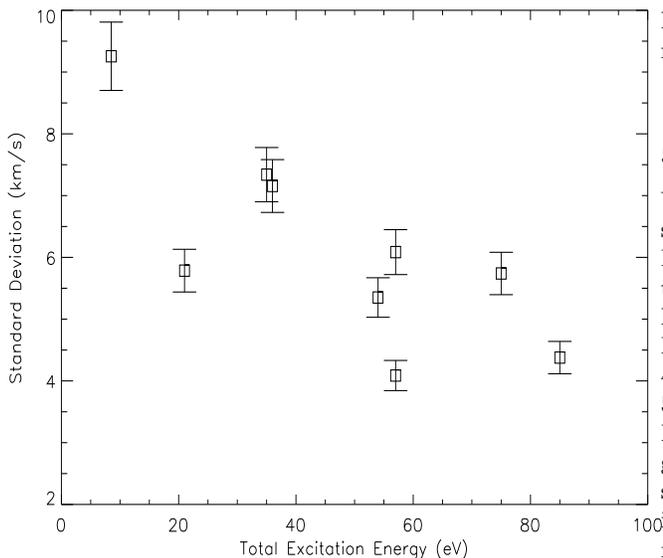,width=9.5cm,height=8.0cm,angle=90}}
\caption[]{Standard deviations of mean radial velocities of
different groups of lines versus TEE. A linear fit ($-$0.049,8.474) 
provides a non-parametric correlation coefficient 0.76.}
\end{figure}

\def\baselinestretch{2}

The radial velocity variations in stars can have instrumental, internal
(atmospheric) and/or external (Keplerian) origin. Instrumental effects
in our measurements are minimized due to the high resolution and high
S/N of the data presented. The internal accuracy achieved for the
wavelength calibrations is of the order 1.5 ${\rm km}~{\rm s}^{-1}$ as
derived from the scatter of measured radial velocities of interstellar 
and telluric lines in the spectra. Note that all former studies 
of HD 188209 (except
for five spectra obtained by Fullerton et al. 1996) were based on 
the photographic spectra. Atmospheric pulsations of early-type stars
have been the subject of extensive studies (e.g. Burki 1978; Bohannan
\& Garmany 1978, Kaufer et al. 1997, Fullerton et al. 1996) 
and have been reviewed by Baade (1988, 1998). 
Gies (1987) has compared the velocity distribution and binary frequency
among 195 Galactic O-type stars (cluster and association, field and runaway)
and found a deficiency of spectroscopic binaries among field stars 
(and especially among the runaway).
The most comprehensive radial velocity studies of OB and BA supergiants
to date are those of Fullerton et al. (1996) and Kaufer et al. (1996; 1997). 
Radial pulsation periods with P$\leq$5$^{\rm d}$ have been predicted
for O-type supergiants (Burki 1978; de Jager 1980; Levy et al. 1984). The 
presence of pulsations or random motions in stellar atmospheres results in
complex velocity curves for different spectral lines due to 
stratification effects (Abt 1957). This is in contrast to Keplerian
motions where all the lines vary synchronously with time (Ebbets 1979;
Garmany et al. 1980). However, in many cases the amplitude 
of the RV variations does not exceed 25--30~${\rm km}~{\rm s}^{-1}$,
and it is very difficult to distinguish whether these variations
are of an internal or an external nature. Additional difficulty comes
from the possible presence of non-radial pulsations (NRP). Variable profiles
(LPVs) have been detected in many narrow-line supergiants (e.g. Baade 1988; 
Kaufer et al. 1997; Fullerton et al. 1996) and in some cases they
correspond to the radial velocity variations measured in photographic
spectra. However, in contrast to broad-line supergiants, clear evidence
of NRPs in narrow- and intermediate-line O-type supergiants has not 
yet been found. Although LPVs are a common occurrence among the O-type
stars, some of them (like HD 34656; Fullerton, Gies \& Bolton 1991)
show a variability which consists of cyclical fluctuations in radial
velocity due to pulsations in a fundamental mode. 

\def\baselinestretch{1}

\begin{figure*}
\vbox{\psfig{file=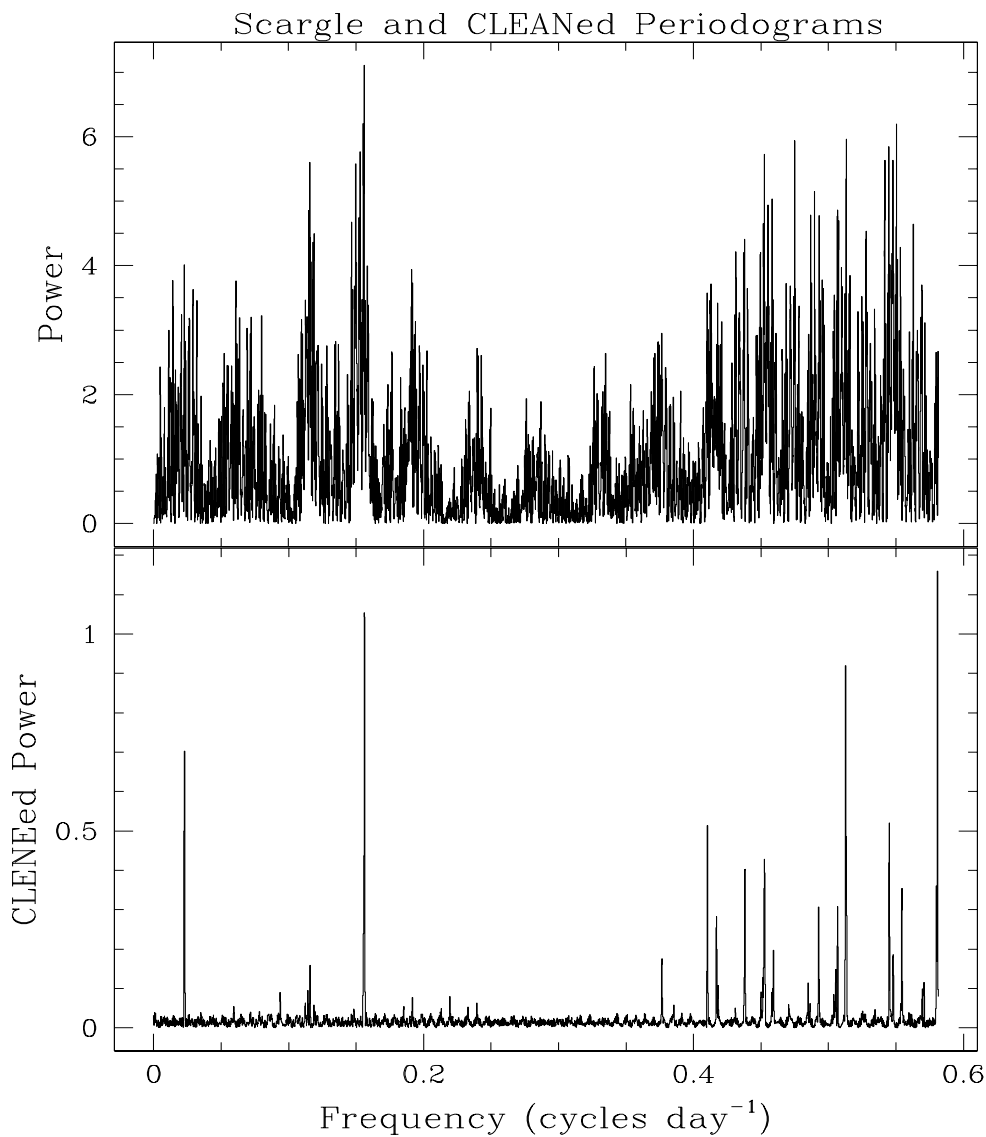,width=17.5cm,height=16.0cm,angle=0}}
\caption[]{Lomb-Scargle (top panel) and CLEAN periodograms of the 
average RV for each date obtained by averaging the RVs 
of all the groups.}
\end{figure*}

\begin{figure*}
\vbox{\psfig{file=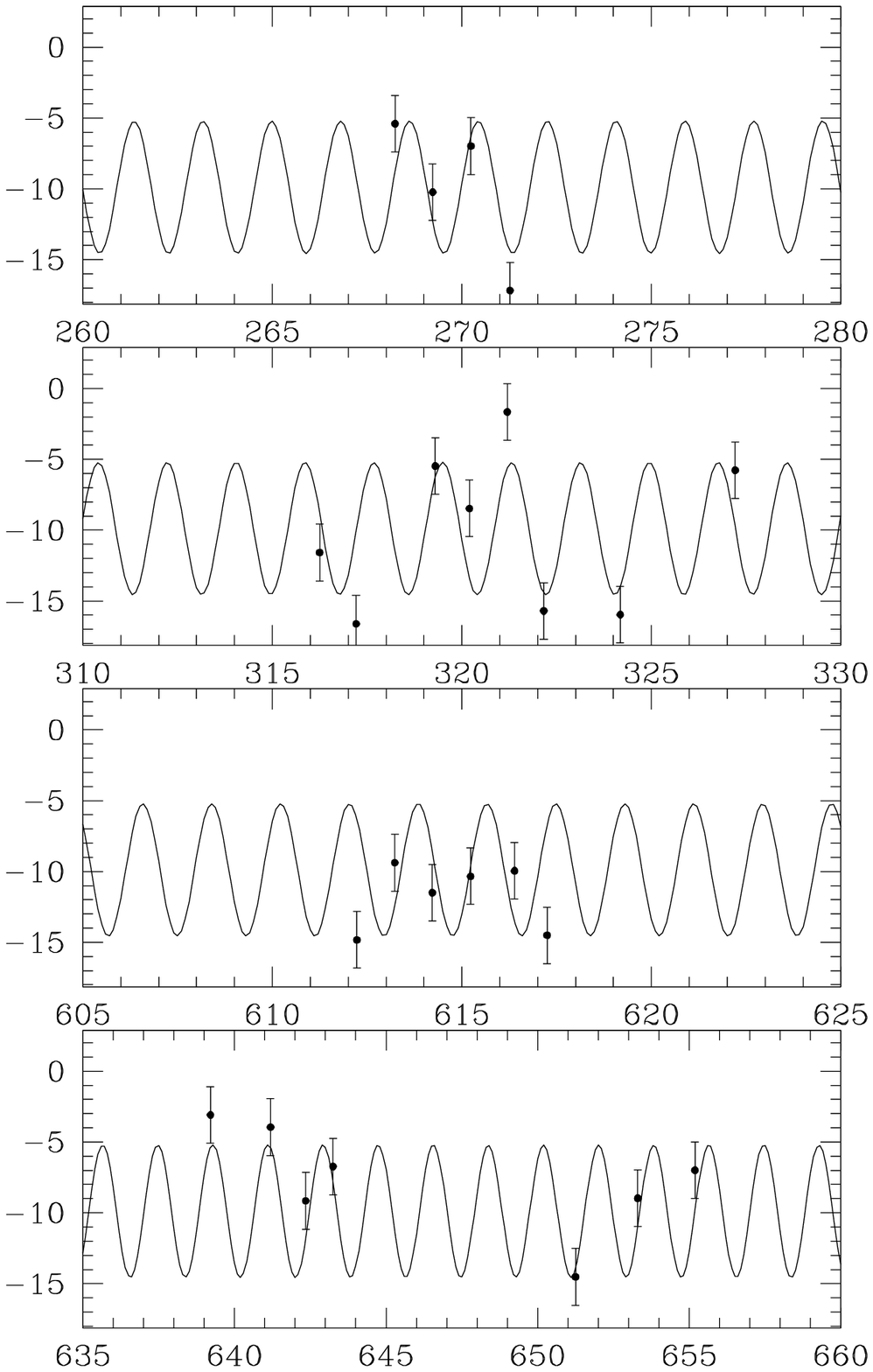,width=15.5cm,height=13.0cm,angle=0}}
\vbox{\psfig{file=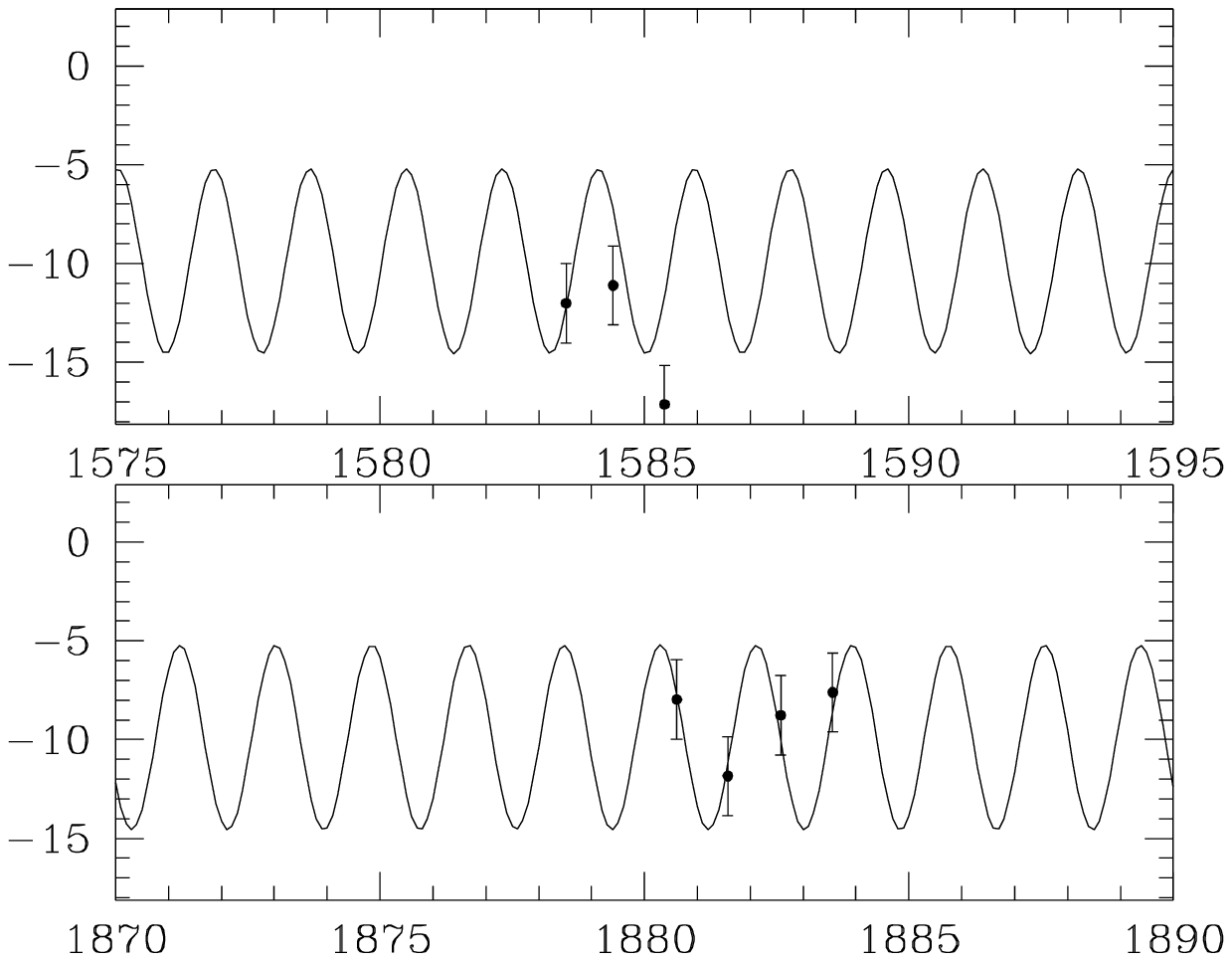,width=15.5cm,height=13.0cm,angle=0}}
\caption[]{The fitting of a sin curve to folded data for a period
6.4 days. The curve is derived for RV=$-$9.8+4.17$\sin$
[2$\pi$(JD-318.629)/6.406].}
\end{figure*}

\def\baselinestretch{2}
 
Careful inspection of our time series showed that all
absorption lines varied in position over the course of a day. 
We have detected a few asymmetric profiles of He lines but there were 
no signatures of moving features. It is however possible that the S/N ratio 
of our data is not high enough to trace LPVs. Small 
distortions in profile shape are typically less than 1\% of the 
continuum strength and a minimum S/N ratio required would be 
at least 300 (Fullerton et al. 1996).

\subsection{The velocity-excitation relationship}

We have selected unblended lines by utilizing theoretical synthetic 
spectra computed for the model atmosphere of HD 188209 and measured RVs 
by fitting a Gaussian to 
the line profile. The following groups of lines have been selected:
\HeI~(4921, 4471, 4713, 5015, 5047, 5876, 6678 \AA), \HeII~(5411,
 4541, 4686 \AA), \SiIV~(4654, 4631 \AA), \CIII~(4650, 4647, 5696 \AA), 
\SiIII~(4567, 4552, 4574 \AA), \CIV~(5801, 5811 \AA), 
\NIII~(4514, 4510, 4518, 4523 \AA), \OII~4661 \AA\ and
\MgII~4481 \AA. 
The average RVs computed for each of these groups of lines are listed 
in Table 2. It is normally assumed that lines of different total excitation
energy (TEE, ionization energy plus excitation energy of the
lower level) form in different layers of the atmosphere (Hutchings 1976). 
However, it is not clear whether the stratification
exists in a dynamically active, pulsating atmosphere. One can 
assume that the time scale of dynamical processes (pulsations,
stochastic motions etc.) is much less than the time necessary for
the establishment of radiative equilibrium. Thus, while pulsating, 
the atmosphere is supposed to pass through a chain of hydrostatic stages.
The assumption that TEE correlates with line formation depths
can be tested as well. We applied our plane--parallel models to
compute formation depths of the line cores of the  \HeI~and \HeII~lines.
It appeared that the core of the strong \HeII~4686 \AA\ line forms
much closer to the surface (at column mass $\sim$ 0.01 g\,cm$^{-2}$) 
than any of the \HeI~lines. However, this was not the case with the
two other \HeII~lines (5411 and 4541 \AA) we used in our study. 
Similar tests have been carried out for other groups of lines 
but assuming an LTE line formation. These exercises suggest 
that before combining lines in different groups and computing 
their average RVs, one has to be sure (of course under the 
assumption that our plane--parallel models are applicable) that their
depths of formation are similar. 

\def\baselinestretch{1}

\begin{table*}
\caption{ Radial velocities of different groups of lines.}
\begin{flushleft}
\begin{tabular}{lllllllllrr}
\hline\noalign{\smallskip}
Date & He\,{\sc i} &  He\,{\sc ii} &  Si\,{\sc iv} & C\,{\sc iii} & Si\,{\sc iii} & C\,{\sc iv} &
N\,{\sc iii} & O\,{\sc ii} 4661 \AA\ & Mg\,{\sc ii} 4481 \AA\ \\
TEE (eV) &  21 & 75 & 57 & 54 & 35 & 85 & 57 & 36 & 8.5 \\
\noalign{\smallskip}
\hline\noalign{\smallskip}
2449176.36 &  $-$12.3 &  $-$11.2 & $-$6.3 & $-$8.1 &  $-$15.9 &  $-$13.2 & $-$8.4 &
$-$26.4 & $-$19.4  \\
2449268.24 &  $-$5.9 &  $-$1.4 &  6.9   & $-$4.9 &  $-$13.8 &  $-$2.9 & $-$5.6 & $-$18.8 & $-$1.3   \\
2449269.23 &  $-$10.9 &  $-$2.1 & $-$0.3 & $-$11.7 &  $-$15.1 &  $-$4.8 & $-$8.0 & $-$24.1 & $-$15.3  \\
2449270.24 &  $-$10.3 &  $-$8.1 &  5.5 & $-$7.5   &  $-$11.7 &  $-$0.1 & $-$6.8 & $-$12.8 & $-$11.6  \\
2449271.27 &  $-$15.7 &  $-$16.5 & $-$7.6 & $-$17.4 &  $-$35.5 &  $-$16.9 & $-$12.3 &
$-$18.2 & $-$14.9  \\
2449294.18 &~   7.9   &~   11.1   &~  6.4 & $-$0.7   &  $-$4.6 &  $-$1.5 &~  2.4   & $-$4.4 & $-$0.6  \\
2449316.25 &  $-$17.3 &  $-$6.5 & $-$1.6 & $-$13.7 &  $-$25.3 &  $-$0.7 & $-$10.8 &
$-$22.7 & $-$4.3  \\
2449317.21 &  $-$17.2 &  $-$12.9 & $-$13.5 & $-$10.4 &  $-$22.5 &  $-$14.7 & $-$12.2 &
$-$24.9 & $-$20.9  \\
2449319.30 &  $-$4.5 &  $-$1.6 &~  3.3   &~  3.1   &  $-$17.1 &  $-$6.1 & $-$5.6 & $-$15.7 & $-$5.4  \\
2449320.20 &  $-$8.5 &~   2.4   &~  1.5   & $-$1.4 &  $-$23.0 &  $-$6.0 & $-$1.3 & $-$27.6 & $-$12.2  \\
2449321.20 &  $-$2.6 &~   2.7   &~  8.4   &~  1.2   &  $-$3.9 &  $-$0.2 &~  0.1   &
$-$14.9 & $-$6.5  \\
2449322.16 &  $-$16.0 &  $-$7.5 & $-$3.5 & $-$9.1 &  $-$16.3 &  $-$11.9 & $-$11.5 &
$-$28.5 & $-$38.1  \\
2449324.18 &  $-$14.8 &  $-$5.2 & $-$7.1 & $-$9.9 &  $-$20.9 &  $-$5.8 & $-$10.6 &
$-$33.5 & $-$36.2  \\
2449327.21 &  $-$1.3 &  $-$1.6 &~  8.6   & $-$3.7 &~   1.4   &~   1.6   & $-$3.7 & $-$20.5 & $-$35.8  \\
2449612.23 &  $-$15.3 &  $-$11.1 & $-$13.9 & $-$11.7 &  $-$22.7 &  $-$11.5 & $-$11.9 &
$-$17.2 & $-$18.8  \\
2449613.23 &  $-$10.6 &  $-$6.2 & $-$7.9 & $-$8.6 &  $-$12.7 &  $-$5.6 & $-$6.2 & $-$15.8 & $-$11.0  \\
2449614.22 &  $-$14.5 &  $-$8.7 & $-$8.2 & $-$10.4 &  $-$19.6 &  $-$6.6 & $-$8.5 &
$-$16.9 & $-$9.6 \\
2449615.23 &  $-$9.2 &  $-$4.7 & $-$1.6 & $-$12.3 &  $-$16.7 &  $-$4.9 & $-$6.4 & $-$15.7 & $-$22.9  \\
2449616.39 &  $-$10.2 &  $-$6.4 & $-$5.4 & $-$8.7 &  $-$16.2 &  $-$4.6 & $-$5.2 & $-$14.2 & $-$20.0  \\
2449617.25 &  $-$15.6 &  $-$7.6 & $-$9.4 & $-$11.5 &  $-$23.8 &  $-$10.4 & $-$10.7 &
$-$23.2 & $-$18.2  \\
2449639.21 &  $-$0.5 &~   0.3   &~  2.3   & $-$0.4 &  $-$8.7 &  $-$1.4 &~  0.2   & $-$13.9 & $-$6.5  \\
2449641.19 &  $-$2.0 &~   3.7   & $-$1.1 & $-$4.1 &  $-$6.3 &  $-$4.1 & $-$3.4 & $-$10.1 & $-$9.5 \\
2449642.35 &  $-$9.7 &  $-$3.1 & $-$3.1 & $-$5.9 &  $-$17.2 &  $-$7.1 & $-$5.9 & $-$16.3 & $-$14.9  \\
2449643.25 &  $-$6.1 &  $-$2.7 & $-$1.2 & $-$6.9 &  $-$14.9 &  $-$1.6 & $-$5.0 & $-$6.4 & $-$18.0  \\
2449651.25 &  $-$13.4 &  $-$13.3 & $-$8.3 & $-$12.3 &  $-$19.9 &  $-$10.5 & $-$11.8 &
$-$23.1 & $-$18.3  \\
2449653.30 &  $-$6.4 &  $-$0.2 & $-$5.5 & $-$6.4 &  $-$9.9 &  $-$8.5 & $-$7.2 & $-$25.0 & $-$11.5  \\
2449655.19 &  $-$8.2 &  $-$4.3 & $-$3.6 & $-$7.5 &  $-$3.4 &  $-$3.3 & $-$7.8 & $-$15.7 & $-$9.7   \\
2450583.52 &  $-$11.6 &  $-$7.5 & $-$7.0 & $-$16.0 &  $-$15.6 &  $-$9.1 & $-$11.3 &
$-$16.4 & $-$13.9  \\
2450584.41 &  $-$13.4 &  $-$5.8 & $-$10.8 & $-$11.6 &  $-$15.9 &  $-$5.6 & $-$11.4 & $-$19.9 & $-$15.3  \\
2450585.38 &  $-$19.2 &  $-$16.9 & $-$13.9 & $-$14.6 &  $-$24.9 &  $-$8.9 & $-$14.2 & $-$25.3 & $-$31.4  \\
2450691.25 &  $-$7.6 &  $-$8.6 & $-$3.2 &~  5.6   &  $-$8.9 &  $-$7.1 & $-$4.8 &~  1.6   & $-$5.9 \\
2450880.61 &  $-$6.5 &  $-$4.6 & $-$3.5 & $-$7.8 &  $-$11.1 &  $-$5.4 & $-$7.8 & $-$8.8 & $-$17.9  \\
2450881.58 &  $-$14.4 &  $-$9.6 & $-$7.3 & $-$13.1 &  $-$14.2 &  $-$6.2 & $-$12.1 & $-$17.3 & $-$23.0  \\
2450882.58 &  $-$11.0 &  $-$4.3 & $-$4.2 & $-$7.7 &  $-$13.6 &  $-$1.2 & $-$7.3 & $-$17.0 & $-$13.3  \\
2450883.56 &  $-$9.1 &  $-$5.2 & $-$4.8 & $-$4.6 &  $-$12.9 &  $-$2.9 & $-$2.7 & $-$15.3 & $-$11.4  \\
Mean~ &  $-$9.6 & $-$5.1 & $-$3.4 & $-$7.7 & $-$15.2 & $-$6.0 & $-$7.3 & $-$17.8 & $-$15.5 \\
Stdv  &~  5.8 &~ 5.7 &~  6.1 &~ 5.3 &~ 7.3 &~ 4.4 &~ 4.1 &~ 7.2 &~ 9.3 \\
\noalign{\smallskip}
\hline
\end{tabular}
\end{flushleft}
\end{table*}

\begin{figure*}
\vbox{\psfig{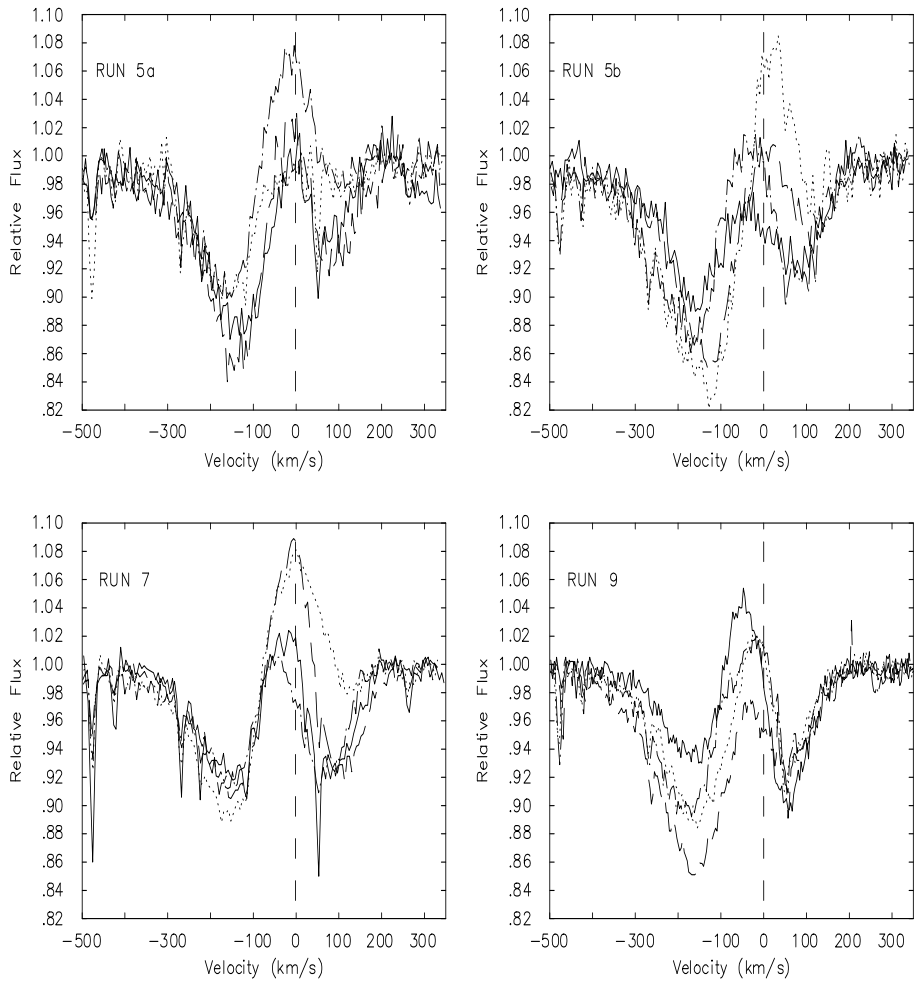}}
\caption[]{The variety of \Ha~profiles observed in different runs. 
A vertical dashed line indicates zero velocity.}
\end{figure*}

\def\baselinestretch{2}

The velocity--excitation relationship found for many hot 
supergiants (Hutchings 1976) exists also in HD 188209.
In Figs. 5 and 6 we present plots of mean RV versus TEE 
and standard deviations of the mean RV versus TEE for 
the different groups of lines computed for all dates (the last 
two lines in Table 2). These plots exclude a pure Keplerian
motion as the only cause of the RV variations. Pulsations and
stochastic motions (intrinsic wind variations caused by some
hydrodynamical instabilities) in the wind can bring to the 
RV variations as well. If we suppose that stochastic 
variations are not important, then the existence of a standard 
deviation-TEE relationship would suggest that deeper layers
in the atmosphere pulsate with smaller amplitudes. The 
amplitude of pulsations increases when approaching
to the surface.

\subsection{The periodicity of the radial velocity variations}

The period search was carried out with help of 
the {\sc period} package (Dhillon \& Privett 1997)
of the {\sc starlink} software. The following strategy was applied when
looking for a periodic signal in the RV curves of different groups of lines.
Due to the large gaps (especially between runs 5 and 6) in our observations,
we first decided to study each of the runs 5, 6 and 7 separately.   
The {\sc clean} algorithm (Roberts et al. 1987) was employed to cover a 
space of loop gains from 0.2 to 0.6 and the number of iterations from 
10 to few hundreds. The convergence of the periodograms was achieved for 
the majority of groups of lines in all three runs. 
The mean frequency suggested by most of the groups in 
all three runs is 0.44$\pm$0.05 days$^{-1}$ (2.24 days). However, this 
period is very close to the Nyquist frequency 
(1/(2$\times$Smallest Data Interval)) of the data and might be
misleading. 
  
We have also looked for periodic signals in the combined data of all 
groups obtained in all runs (Table 2). The maximum and minimum 
frequencies were set to 100 and 0, respectively. A {\sc clean} analysis of 
the time series of the majority of groups revealed a frequency 
of 0.51$\pm$0.1 days$^{-1}$ (1.95 days). The gain factor was 0.1 at 
the first iteration, then was decreased by 15-20 
iteration until stabilization. The average RV for each date 
obtained by averaging the RVs  of all the groups revealed a 
frequency of 0.47$\pm$0.12 days$^{-1}$ (2.1 days). 
Again, both frequencies are very close to the Nyquist frequency and we
should discard them. We must point out that our periodograms did not show
any peaks at frequencies smaller than  0.4 days$^{-1}$. The next strongest 
peak which appeared in our periodograms was near 0.156$\pm$0.15 days$^{-1}$
(6.4 days). Clearly this period is not affected by sampling.
We have also analysed the RV data using the Lomb-Scargle method
(Lomb 1976, Scargle 1982) which allows to compute statistical
probability of peaks in periodograms. To ensure reliable significance
values, the minimum number of permutations was set 100. The probability that
the period is not equal to 6.4 days was always less than 30 $\%$. The
peak at 6.4 days appears in all periodograms but given its significance
value, we cannot definitely rule out its non-physical nature. 
In Figs. 7 and 8 we show the  {\sc clean} and the {\sc lomb-scargle}
periodograms and the fitting of a sin curve to folded data, respectively.  

\def\baselinestretch{1}

\begin{figure*}
\vbox{\psfig{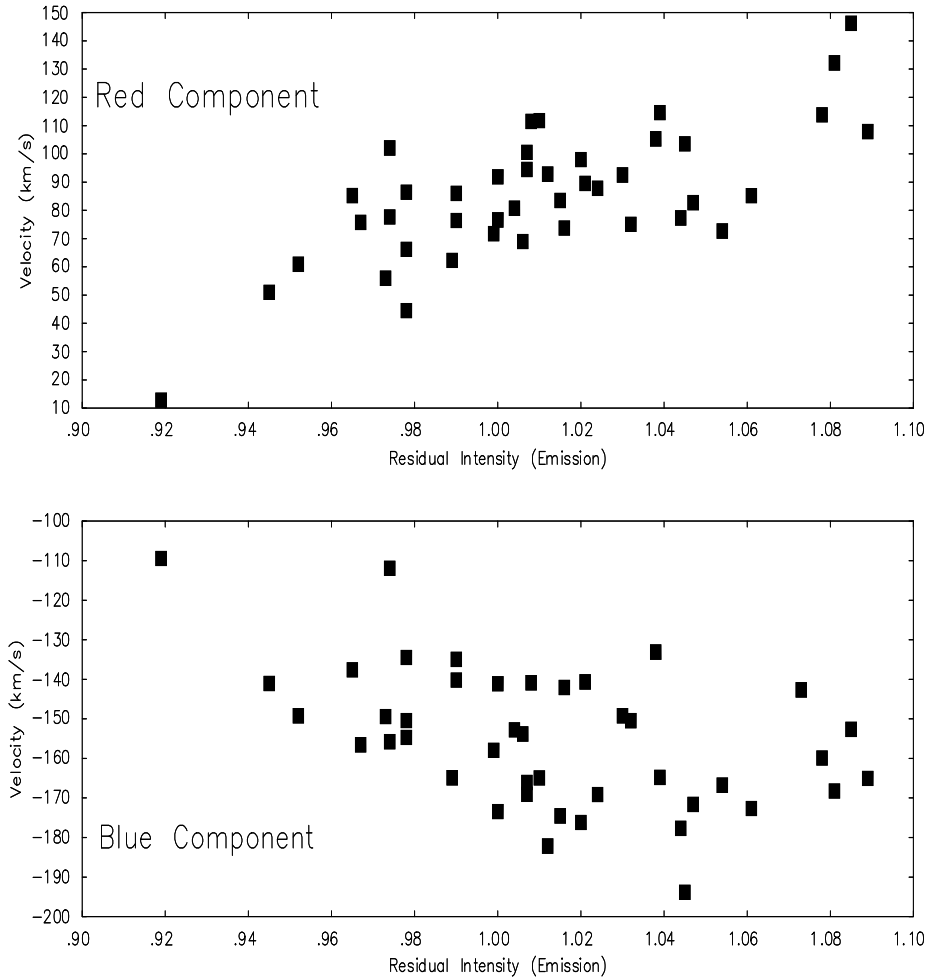}}
\caption[]{The radial velocities of blue and red absorption components
versus residual intensity of the central emission.}
\end{figure*}

\def\baselinestretch{2}

\section{Variability of H$\alpha$}

All hot supergiants have variable \Ha~profiles in their
spectra (Rosendahl 1973). The shape of the \Ha~may vary from P Cyg to 
inverse P Cyg, double-peaked, pure absorption and/or emission 
(Ebbets 1982) with typical time-scales of the order of days. 
The nature of this variability is not yet understood.  
The existence of variable asymmetric outflows/infalls of
matter and some corotating structures related to surface 
inhomogenities and possible magnetic fields have been proposed for 
BA-type (Kaufer et al. 1996) and O-type (Fullerton et al. 1996; 
Kaper et al. 1997)  supergiants.
In addition, there have been detailed studies of the rotating
giant loop in $\beta$ Orionis (Israelian, Chentsov \& Musaev 1997) 
and the corotating spiral structures in HD 64760 and HD 93521 
(Howarth et al. 1998; Fullerton et al. 1997). It is of course 
very difficult to distinguish binary systems from single 
stars without understanding the nature of \Ha~variability. 
As Thaller (1997) suggests, the \Ha~can suffer some peculiar
variability due to the colliding winds in a binary system.  

The time evolution of \Ha~profiles in three different runs is shown 
in Figure 9. The average \Ha~profile consists of three components, 
a central emission accompanied by blue and red absorptions.
We have not observed a single \Ha~profile without a central 
reversal. The emission is not always centered exactly on the rest 
wavelength but is varying. It may approach the continuum level, 
go above it and decrease rapidly in strength. Apparently
the time-scale of the \Ha~variability is at least one day. 
The 5$^{\rm th}$ run has been divided into two parts
(runs 5a \& 5b) with four successive nights in each. 
The \Ha~variability is observed over a wide range from 
about $-$400 to 200~${\rm km}~{\rm s}^{-1}$. 
 
We have already seen in Section 4.2 that our spherical unified 
models can account for the central reversal (or at least set
upper and lower limits of the mass-loss). Thus, we know that
the central emission forms in the expanding envelope and accounts
for the filling-in effect observed in \Ha, \Hb~and some other lines.
The filling-in effect observed in \Hb~ correlated perfectly  with
the strength of the \Ha~central reversal. The \Ha~ wings 
originate deep in the atmosphere, whereas the central reversal comes
from the thin layers of the envelope. One cannot use the term
``underlying photospheric absorption line'' since the central
reversal is not emitted by a detached layer far from the 
photosphere. It is important to stress that we deal with a
$single$ line formed in a $unified$ model atmosphere. 
A variable amount of incipient emission can be due to density 
(or radius) variations in the outer
atmosphere. However, these variations are not expected to produce
an asymmetry as long as we deal with spherically symmetric
mass-loss. Our observations indicate that the velocity of the
central emission varies as well. We $do$ expect RV 
variations in the central emission (the upper atmosphere) since we know that
the RVs of all absorption lines vary. The amplitude of the RV variations 
of \MgII~4481 \AA\ can reach 30$~{\rm km}~{\rm s}^{-1}$ (Table 2) and
40$~{\rm km}~{\rm s}^{-1}$ in \Hb. Thus, it is not unusual for the
amplitude of RV variations of the \Ha~central emission to reach 
50$~{\rm km}~{\rm s}^{-1}$. 
A period analysis of the RV curves of the central emission resulted in the 
detection of quasi-periodic variability with a frequency
0.42$\pm$0.14 d$^{-1}$ (2.35 days). It turns out that the RV 
curves of all groups of lines vary in phase. However, the RV curve 
of the \Ha~was shifted half phase relative to all groups.    

We have measured the RVs of the blue and red absorption components
of the \Ha~and plotted them against the residual intensity of the central
emission reversal. This plot (Fig. 10) shows a correlation
with a large scatter due to the RV variations of the central reversal.
This kind of correlation can be expected when the central reversal
is moving up and down relative to a local continuum. We found similar
correlations between the RV of the central emission and the residual
intensities of the red and blue absorptions.
Apparently all the changes observed in the
\Ha~wings at velocities $v \leq -$ 100$~{\rm km}~{\rm s}^{-1}$ and 
$v \geq$ +100$~{\rm km}~{\rm s}^{-1}$ are due to the variations
of the central reversal. We do not anticipate such  large
RV variations deep in the atmosphere where these wings are formed. 
The conclusion is that the overall shape of the \Ha~is determined by the 
central emission.  

We have performed a period search of the integrated equivalent width 
(EW) of the \Ha~data set and found a maximum in power at frequency 
0.22 day$^{-1}$. Figures 11 and 12 show the phase diagram for the
period 4.41 days and a grey-scale representation of the phase 
spectrum, respectively. A phase spectrum represents a two-dimensional
case of the {\sc clean} algorithm where each velocity bin of the 
\Ha~is treated as a time series of the \Ha~intensity. 

\def\baselinestretch{1}

\begin{figure}
\vbox{\psfig{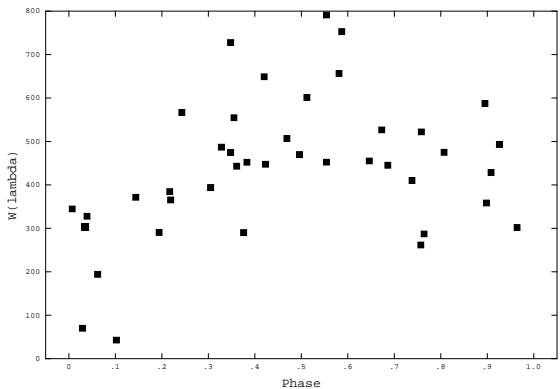}}
\caption[]{The phase diagram of the EWs of \Ha.}
\end{figure}

\begin{figure}
\label{dyn}
\psfig{figure=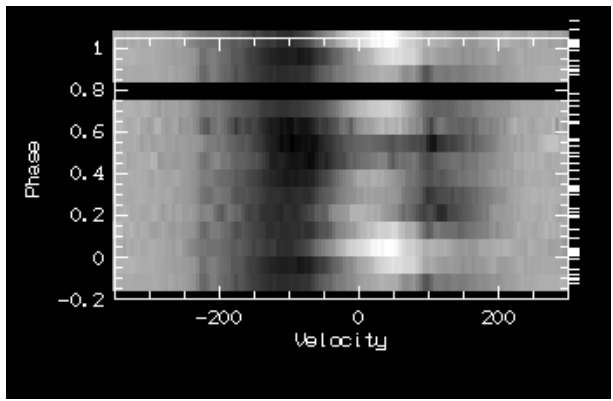,width=8.0cm}
\caption[]{Dynamical phase spectrum of \Ha.}
\end{figure}

\def\baselinestretch{2}

\section{Discussion}

Our target belongs to the group of stars for which the existence of
a compact companion has been proposed in the literature. 
The task of disproving or confirming the  binary nature of the system 
can be tackled only if sufficiently accurate analysed 
observational data are available. In this paper we used state-of-the-art 
models of atmospheres to determine the fundamental parameters of
HD 188209.

To establish the presence of a possible companion we have studied
the RVs of absorption lines by combining them in different groups. 
Fourier analysis based on the iterative {\sc clean} 
algorithm was used to search for periodic variability. 
Unfortunately the time coverage of runs 1--4 and 8--9 was too
sparse to set constraints on their time-dependent behaviour. 
For this reason we first analysed a few runs separately and
then utilized the {\sc clean} algorithm to search for periods in
a whole data set. The highest peak in the Fourier power spectrum 
was centered near the frequency 0.156 day$^{-1}$ (6.4 days). 

The 6.4 days period can be due to the binary nature of the system
if one assumes very small and unlikely values for the mass ratio (q$\le$0.1). 
Taking the values derived in this article (\M = 16.6, \R =20.9)
and assuming q=0.1, we obtain for the Roche radius  and
for the major semi-axis of the binary orbit 
15 \Rsun~and 25 \Rsun, respectively. This simple 
estimate shows that an O supergiant can hardly fit 
within the orbit because its Roche radius would be less than 
the stellar radius.  Even if it would fit, the tides in such a
tight binary would be very strong making the star 
to speed up quickly until the rotation period matches the orbit. 
Even if we assume that the system is very young, it's 
hard to explain that the putative orbital period is 2 times 
shorter than the rotational period (13 days). 
The second difficulty with the binary interpretation comes
from the variability of H$\alpha$. A TVS (temporal variance spectrum,
showing the extent and distribution of statistically significant 
profile variability) has been computed recently 
(Baade 1998b, Kaper et al. 1998) for 15 spectroscopic binaries and
it was found that all they show a characteristic double-peaked 
profile. This is due to two H$\alpha$ absorption/emission profiles
moving in a composite spectra. In our case the H$\alpha$ profile 
is splitted because of the central emission coming from the lower
wind. The last argument comes from the clear relations between
excitation energies and radial velocity amplitude and excitation energy
and mean radial velocities and from the model atmosphere atmosphere 
calculations. The latter is a good discriminant between
internal variations (pulsations \& wind instabilities) and 
Keplerian motions. In a binary system one would expect all lines to
have the same amplitude independent on their TEE. 

It has been known for a long time (Abt 1957) that the quasi-periodicity 
in hot supergiants might be ascribed to radial pulsations. 
A simple relation (Burki 1978; de Jager 1980) can be used to estimate 
the period of radial pulsation,
\begin{equation}
\log P_{\rm fund} = 10.93 - 0.5\log (M/M_{\odot}) - 0.38M_{\rm bol} - 3 \log 
T_{\rm eff}
\end{equation}
Using the values of parameters obtained in Section 4 we arrive
at $P_{\rm fund}$=1.75 d. 
Note that the form of the relation (1)
depends on the stellar evolutionary models and the input parameters;
both are subject to large errors. In particular, note that we
found no large differences in the parameters determined with 
plane--parallel and unified model atmospheres. Nevertheless, Levy et al. (1984)
have pointed out that periods a factor of 1.5 longer than the
corresponding periods of the radial pulsations can be ascribed to
non-radial pulsations. This means that a factor of two difference 
between the evolutionary and the spectroscopic masses can easily
result in the mis-identification of the pulsating mode.
Another difficulty has been pointed by the referee of the article.
A more sophisticated approach shows (Unno et al. 1979) that f-mode 
pulsation (which is the lowest-frequency mode supported by radial 
pulsation) periods are about 10 times larger than the one suggested
by a period-luminosity relation.  In any case, the theoretical 
period of 1.75 days is very close to the Nyquist frequency of our
data which means that we have a little chance to identify it in 
our data set even if it exists.

Our data not allow to distinguish between pulsations and
stochastic variations of the stellar wind. It is also  
quite possible that we have a combination of both effects.  
 
Note that the projected rotational period 
of this star ($\sim$ 13 d) is much longer than any of the quasi-periods 
found in this paper (but of course our runs do not cover a whole
rotation cycle). The surface features (if any) will always be
visible on the projected disc of the star independently of the inclination
angle. Thus, any periods due to the rotation of these features must 
correspond directly to the rotation period. We do not find any
peaks in the power spectra at $\sim$13 d and this leads us to discard
rotational modulation as a possible explanation of the RV variations 
reported here.   

The quality and the sampling of our data do not
allow a careful study of the line asymmetries, moving components
(if later exists) and/or long-term spectroscopic variability to be made.
It is quite possible that the non-sinusoidal character of the RV curve
for 6.4 days period (Fig 8)
is caused by some disturbances due to the NRPs and/or moving features
in the profiles plus any stochastic instabilities of the wind. 
New monitoring with much higher S/N may allow NRPs, multimode 
pulsations and clearly separate a sinusoidal curve of the radial 
pulsations to be revealed. However, we found convincing 
evidence that the atmospheric motions cannot be ascribed solely 
to Keplerian motions and probably are not of a binary origin.

\section*{Acknowledgments}
 
We thank D.~Baade, O.~Pols and Pablo Rodriguez for their useful comments. 
A.~G. thanks the Canadian Astronomical Society for the travel grant 
to SAO,  and I.~Bikmaev for helpful discussions. We wish to thank 
the anonymous referee for his careful reading of the manuscript and 
several constructive suggestions.

\end{document}